\documentclass[preprint,12pt]{elsarticle}
\biboptions{sort&compress}

\usepackage[left=2cm,right=2cm,bottom=2.5cm,top=2.5cm]{geometry}
\usepackage{epsfig}         
\usepackage{cuted}
\usepackage{amsmath}
\usepackage{amssymb}
\usepackage{float}
\usepackage{xcolor}         
\usepackage[normalem]{ulem} 
\usepackage{multirow}
\usepackage{multicol}
\usepackage{subcaption}
\usepackage{lscape}
\usepackage{bm}
\usepackage{verbatim}
\usepackage{graphicx}
\usepackage{psfrag}
\usepackage{hyperref} 
\usepackage[export]{adjustbox}
\usepackage{lineno}
\usepackage{soul}
\usepackage{setspace}

\definecolor{b}{rgb}{0, 0.4470, 0.7410}
\definecolor{r}{rgb}{1, 0.3010, 0.3010}
\definecolor{w}{rgb}{1, 1, 1}

\begin{document}

\begin{frontmatter}

\title{Design and characterization of cochlea-inspired tonotopic resonators}

\author{Vin\'icius F. Dal Poggetto\fnref{label1}}
\author{Federico Bosia\fnref{label2}}
\author{David Urban\fnref{label3}}
\author{Jan Torgensen\fnref{label4}}
\author{Nicola M. Pugno\fnref{label1,label5}}
\author{Antonio S. Gliozzi\fnref{label2}\corref{cor1}}\ead{antonio.gliozzi@polito.it}

\fntext[label1]{Laboratory for Bio-inspired, Bionic, Nano, Meta Materials \& Mechanics, Department of Civil, Environmental and Mechanical Engineering, University of Trento, 38123 Trento, Italy}
\fntext[label2]{DISAT, Politecnico di Torino, 10129 Torino, Italy}

\fntext[label3]{Department of Electronic Systems, Norwegian University of Science and Technology, NO-7491 Trondheim, Norway}

\fntext[label4]{State Materials Testing Laboratory for Mechanical Engineering, Technical University of Munich, Boltzmannstrasse 15, Garching, Germany}

\fntext[label5]{School of Engineering and Materials Science, Queen Mary University of London, Mile End Road, London E1 4NS, United Kingdom}

\cortext[cor1]{Corresponding authors}

\begin{abstract}
\textit{The cochlea has long been the subject of investigation in various research fields due to its intriguing spiral architecture and unique sensing characteristics. One of its most interesting features is the ability to sense acoustic waves at different spatial locations, based on their frequency content. In this work, we propose a novel design for a tonotopic resonator, based on a cochlea-inspired spiral.
The resulting structure was subjected to an optimization process to exhibit out-of-plane vibration modes with mean out-of-plane displacement maxima distributed along its centerline spanning nearly a two-decade frequency range.
Numerical simulations are performed to demonstrate the concept, which is also confirmed experimentally on a 3D printed structure. The obtained frequency-dependent distribution is shown to be a viable source of information for the discrimination of signals with various frequency components. The harnessed tonotopic features can be used as a fundamental principle to design structures with applications in areas such as non-destructive testing and vibration attenuation.}
\end{abstract}

\begin{keyword}
Bio-inspired metamaterial; Cochlea; Tonotopy; Resonators.
\end{keyword}

\end{frontmatter}

\section{Introduction} \label{introduction}

Among the various sensory organs developed in nature over the course of evolution, the cochlea is certainly one of the most fascinating \cite{dallos2012cochlea}. This coiled structure possesses unique characteristics which allow mammalians to perceive sounds in wide frequency and amplitudes ranges, comprising nearly ten octaves and $120$ decibels \cite{robles2001mechanics}. As one of the many examples of naturally-occurring spiral structures \cite{livio2008golden}, the cochlea can be approximately modeled by considering a logarithmic spiral description, first presented by D\"{u}rer \cite{durer1525underweysung}, which can be used to investigate the influence of geometrical features on its frequency-dependent characteristics \cite{manoussaki2008influence}. Beyond the obvious advantages offered by structural coiling in terms of packing, the cochlea also possesses tonotopic characteristics, i.e., where the intensity of the mechanotransduction responsible for converting sound energy into neural impulses varies with the location of peak excitation \cite{lemasurier2005hair}. Such space-dependent detection of mechanical signals based on their frequency content provides unique frequency detection attributes that are essential for hearing \cite{lighthill1991biomechanics}.

The design concepts present in the cochlea are also of considerable interest in the field of bio-inspired metamaterials, usually employing locally resonant structures able to mimick the tonotopy observed in the cochlea \cite{ma2016cochlear,rupin2019mimicking}. The use of locally resonant structures is widespread in the context of wave manipulation, ever since the seminal work presented by Liu et al. \cite{liu2000locally} opened new possibilities for the development of locally resonant phononic crystals \cite{kushwaha1994theory,laude2020phononic}, with one- \cite{xiao2013theoretical,xiao2013flexural,dal2019optimization}, two- \cite{xiao2012flexural,miranda2019flexural,poggetto2021flexural}, and three-dimensional applications \cite{mitchell2014metaconcrete,witarto2019three,dal2020elastic}. These structures present as a typical characteristic local resonance band gaps \cite{bloch1929quantenmechanik}, which are frequency bands that strongly attenuate wave propagation and can occur at sub-wavelength dimensions \cite{krushynska2017coupling}. Also, in contrast with the typical band gaps observed in phononic crystals, occurring due to Bragg scattering \cite{gorishnyy2005sound}, local resonance band gaps do not strictly require the periodicity of the medium to emerge, thus avoiding the necessity of using many unit cells to achieve a noticeable effect.

The width of local resonance band gaps, on the other hand, is typically much narrower than Bragg scattering band gaps \cite{krushynska2017coupling}. To overcome this drawback, a possible approach is to combine several elements with distinct resonant frequencies to create a rainbow effect able to encompass a wider frequency range \cite{de2021selective}. The experimental realization of structures able to achieve this effect, however, may be complicated, since it requires the combination of multiple elements \cite{meng20203d}. Approaches considering structures able to separate frequencies along their spatial profile for impinging acoustic waves have been proposed in linear \cite{zhu2013acoustic} and coiled structures \cite{ni2014acoustic}, but to the best of our knowledge, no similar designs were attempted considering elastic waves propagating in a solid medium. Therefore, the current state-of-art lacks the development of a structure which can be used with the same simplicity as that of a tuned resonator \cite{jin2017pillar}, but does not suffer the drawbacks of being restricted to a limited number of resonant frequencies, thus presenting the advantages of the rainbow-like effect which is associated with the utilization of a cochlear geometry.

The numerical and experimental results presented herein demonstrate how a cochlea-inspired tonotopic resonator can be used as a single resonant element with tailored resonant frequencies able to span a wide frequency range, thus expanding the possibilities of wave manipulation and control, with relevant technological use involving mechanical vibrations, such as sensing applications \cite{pennec2019phononic} and energy harvesting \cite{carrara2013metamaterial,qi2016acoustic,chaplain2020topological,de2020graded,de2020experimental,lin2021piezoelectric}.

The paper is organized as follows. In Section \ref{models_methods}, the geometrical model for the development of the cochlea-inspired resonator is presented, along with relevant metrics which are useful to investigate the spatial distribution of energy according to the variation of natural frequencies, i.e., the tonotopic energy profile. The numerical and experimental results illustrating the tonotopic behavior of the proposed curved structure are presented in Section \ref{results}. Concluding remarks are presented in Section \ref{conclusions}.

\section{Models and methods} \label{models_methods}

\subsection{Geometric properties}

\begin{figure}[ht!]
 \centering
 \includegraphics[scale=0.45]{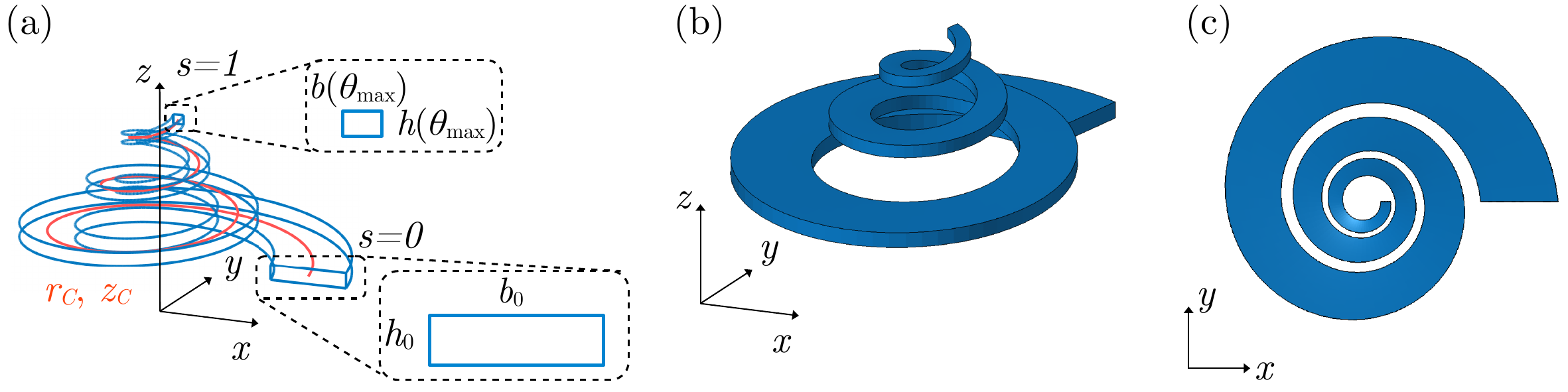}
 \caption{
 Model of the cochlea-inspired curved structure.
 (a) The rectangular cross section ($b(\theta) \times h(\theta)$) is swept along the curve $C$ described in the cylindrical coordinate system as ($r_C, \, \theta, \, z_C)$, creating a variation between the initial cross section ($b_0 \times h_0$) and the final cross section ($b(\theta_{\max}) \times h(\theta_{\max})$).
 An example of a cochlear structure constructed using this approach is shown in 
 (b) perspective and (c) top views, respectively.
 }
 \label{cochlea_geometry}
\end{figure}

The cochlea-inspired curved structure is here modeled by sweeping a rectangular cross section $b(\theta) \times h(\theta)$ (width and height, respectively) along a centerline curve, $C$, described in a cylindrical coordinate system $(r_C, \, \theta, \, z_C)$ (Figure \ref{cochlea_geometry}a). The mass density of the structure is taken as constant. A logarithmic spiral can be written in these coordinates as
\begin{equation} \label{geo_rz}
 r_C(\theta) = r_0 e^{k_r \theta / \theta_{\max} } , \;
 z_C = z_C(\theta), \;
 \theta \in [0, \, \theta_{\max}] \, ,
\end{equation}
where $r_0$ is the initial radius of the spiral, $k_r \in \mathbb{R}^*$, represents the constant polar slope of the spiral, and $\theta_{\max} = n_T 2\pi$ is the maximum angular coordinate for a number $n_T$ of turns of the cochlea.
Planar cochleae are represented by the simplification $z_C(\theta)=0$.
The arc length $l$ can be computed using the line integral over the radius function in Eq. (\ref{geo_rz}) as
\begin{equation}
 l(\theta) = \int_0^{\theta} \, \sqrt{
 r^2_C(\xi) + \bigg( \frac{dr_C(\xi)}{d\xi} \bigg)^2 + \bigg( \frac{dz_C(\xi)}{d\xi} \bigg)^2
  } \, d\xi \, ,
\end{equation}
which, for the case of $z_C = 0$, simplifies to
\begin{equation}
 l(\theta) = \frac{r_0 \theta_{\max}}{k_r} \sqrt{ 1 + \bigg(\frac{k_r}{\theta_{\max}} \bigg)^2 } \, ( e^{k_r \theta/\theta_{\max}} - 1 ) \, ,
\end{equation}
thus leading to the centerline normalized coordinate $s \in [0, \, 1]$, described as
\begin{equation} \label{s}
 s(\theta) = \frac{l(\theta)}{l(\theta_{\max})} = \frac{e^{k_r \theta/\theta_{\max}} - 1}{e^{k_r} - 1} \, .
\end{equation}

Analogously, the cross section dimensions of the cochlea, $b = b(\theta)$ and $h = h(\theta)$, can be written as
\begin{equation} \label{geo_bh_theta}
 b(\theta) = b_0 e^{k_b \theta / \theta_{\max}} , \;
 h(\theta) = h_0 e^{k_h \theta / \theta_{\max}} \, ,
\end{equation}
with $b_0$ and $h_0$ representing, respectively, the initial width and height of the cochlea cross section, and $k_b \in \mathbb{R}$ and $k_h \in \mathbb{R}$ associated with the variation of the width and height, respectively. These equations can also be expressed in terms of the normalized coordinate $s$ using Eq. (\ref{s}):
\begin{equation} \label{geo_bh_s}
 b(s) = b_0 [ 1 + s(e^{kr}-1) ]^{k_b/k_r} , \;
 h(s) = h_0 [ 1 + s(e^{kr}-1) ]^{k_h/k_r} \, ,
\end{equation}
thus indicating the explicit dependence of $b$ ($h$) on the ratio $k_b/k_r$ ($k_h/k_r$), which can be varied to tune the geometric parameters of the spiral.

The parameters $k_b$ and $k_h$ can be defined by specifying the cochlea cross section at $\theta = \theta_{\max}$, i.e., $k_b = \ln(b(\theta_{\max})/b_0)$ and $k_h = \ln(h(\theta_{\max})/h_0)$. An example of a cochlea constructed using this approach is shown in Figure \ref{cochlea_geometry}b.
Also, special care must be taken so that overlapping does not occur (i.e., the cochlea does not self-intersect, see Figure \ref{cochlea_geometry}c).

The volume of the resulting cochlea is calculated as
\begin{equation} \label{geo_volume}
 V = \int \limits_V \, dV
 = \int_0^{\theta_{\max}} \, \int_{z_C - h/2}^{z_C + h/2} \, \int_{r_C - b/2}^{r_C + b/2} \,
 r \, dr \, dz \, d\theta
 = \int_0^{\theta_{\max}} \, r_C(\theta) \, b(\theta) \, h(\theta) \, d\theta
 = r_0 b_0 h_0 \, \alpha \, ,
\end{equation}
where $\alpha = (e^{k_r+k_b+k_h}-1) \theta_{\max} / (k_r+k_b+k_h) $ can be regarded as a dimensionless shape factor.

For simplicity, let us assume a planar cochlea. The gap in the radial direction between two adjacent turns is given by
\begin{equation} \label{geo_gap}
 \Delta_r(\theta) = r_i(\theta) - r_e(\theta+2\pi) \, , 
\end{equation}
where $r_i(\theta) = r_C(\theta) - b(\theta)/2$ and $r_e(\theta) = r_C(\theta) + b(\theta)/2$ are the inner and outer radii of the cochlea, respectively, for the angular coordinate $\theta$.
Although the condition $\Delta_r(\theta)=0$ is theoretically possible, it implicates in surface contact and manufacturing issues.
Thus, the condition of no overlapping is achieved for $\Delta_r(\theta) > 0$, $\theta \in [0, (n_T-1)2\pi]$, which can be related with a minimum gap value ($\Delta_r^{\min}$) as
\begin{equation}
 \Delta_r(\theta) = r_c(\theta) - b(\theta)/2 - (r_c(\theta+2\pi) + b(\theta+2\pi)/2 ) \geq \Delta_r^{\min} \, ,
\end{equation}
which, combined with Eqs. (\ref{geo_rz}) and (\ref{geo_bh_theta}), yields the condition
\begin{equation} \label{geo_r0}
 r_0 e^{k_r \theta/\theta_{\max}} ( 1 - e^{k_r/n_T} )
 - (b_0/2) e^{k_b \theta/\theta_{\max}} ( 1 + e^{k_b/n_T} ) \geq \Delta_r^{\min} \, .
\end{equation}

Thus, for a given fixed volume $V$, geometric parameters $\{ k_r, \, k_b, \, k_h, \, n_T, \, h_0 \}$ and a minimum gap $\Delta_r^{\min}$, Eqs. (\ref{geo_volume}) and (\ref{geo_r0}) can be used to determine $b_0$ and $r_0$.

\subsection{Tonotopic modal profiles}

\begin{figure}[H]
 \centering
 \includegraphics[scale=0.45]{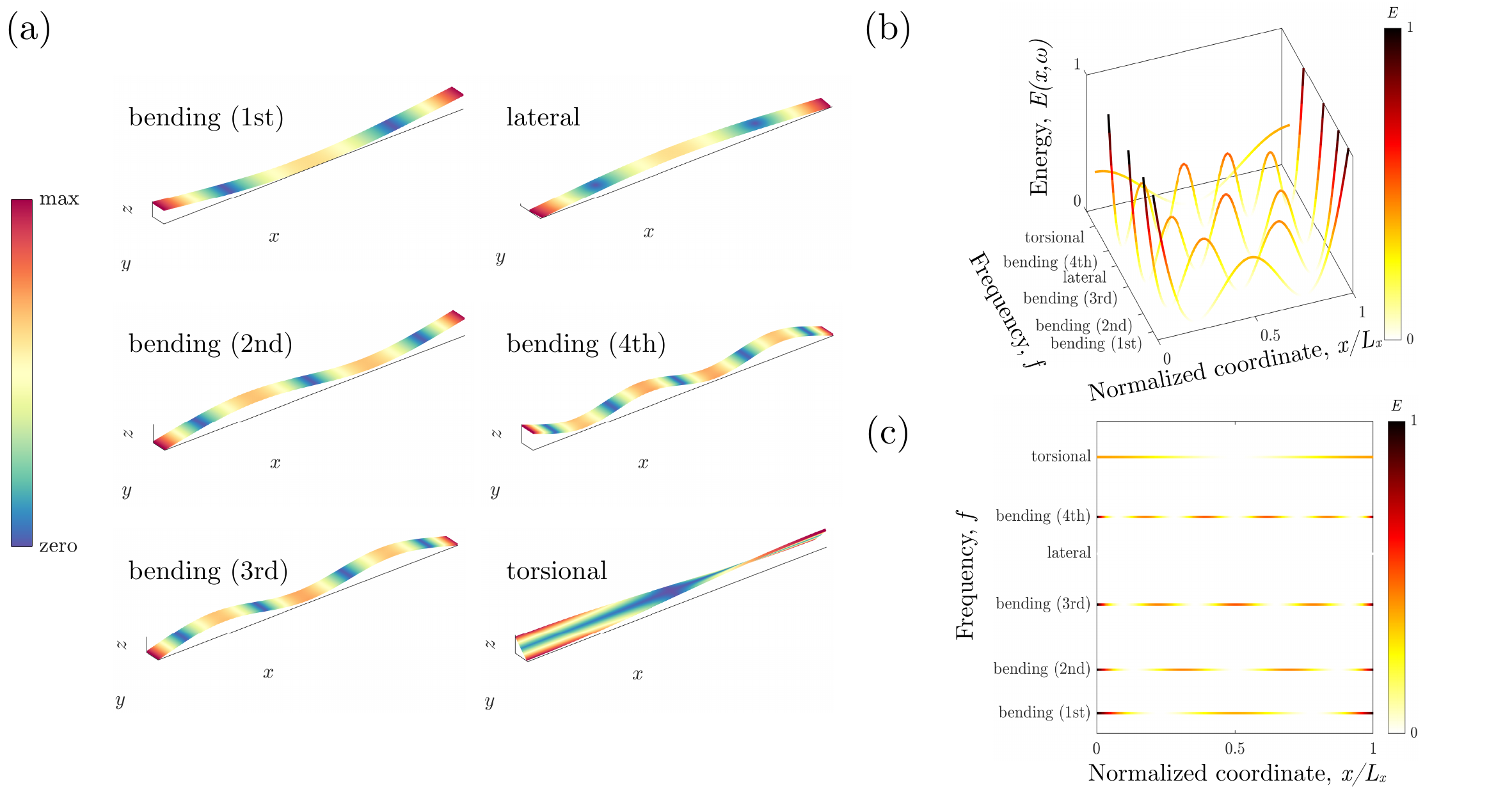}
 \caption{Representations of the tonotopic modal profile for a rectangular prism with dimension $L_x \times L_y \times L_z$ ($L_x \gg L_y \gg L_z$).  (a) The first $6$ eigenmodes, with the color scale representing absolute values of displacements.
 (b) Tonotopic modal profile representation displaying the quadratic out-of-plane displacements mean. (c) Top views of the same plot in (b).}
 \label{tonotopic_frequency_profiles}
\end{figure}

As a reference case, we initially consider a rectangular prism with dimensions $L_x \times L_y \times L_z$ with $L_x \gg L_y \gg L_z$. The resonant frequencies and corresponding modes of vibration of this element can be obtained for a given set of boundary conditions. If one considers, for instance, free boundary conditions for all of its faces, the first $6$ modes (excluding rigid-body motion) are depicted in Figure \ref{tonotopic_frequency_profiles}a, with the color bar representing the absolute value of displacements. At this point, no specific material properties are specified, since the presented concepts are general.

The representation of each separate eigenmode, however, may become increasingly difficult for large sets. Thus, it may be useful to summarize the information contained in each eigenmode in a concise manner.

This can be done, for instance, by considering the eigenmode displacements in a given direction, e.g., out-of-plane displacements ($u_z = u_z(x,y,z,\omega)$), and calculating an associated
quadratic displacement cross-section mean (mean displacement for short), $E$, computed over its transverse directions ($y$ and $z$), written as 
\begin{equation} \label{energy_integral}
 E(x,\omega) =
 \frac{1}{A} \int \limits_A \,  |u_z(x,y,z,\omega)|^2 \, dy \, dz
 = \frac{1}{A} \int \limits_A \,  u_z^*(x,y,z,\omega) \, u_z(x,y,z,\omega) \, dy \, dz \, ,
\end{equation}
where $A$ is the cross section at coordinate $x$, $(\cdot)^*$ denotes the complex conjugate operator and can be calculated for each eigenmode, previously normalized with respect to its largest computed displacement. Out-of-plane direction displacements are chosen due to (i) a biological motivation, since such displacements are analogous to the direction in which the cochlea basilar membrane is coupled with the surrounding fluid, and (ii) an experimental motivation, since out-of-plane velocities can be directly measured in an experimental setup. The quantity $E(x,\omega)$ can be computed for a set of resonant modes of a given frequency range of interest, thus yielding a tonotopic modal profile ($E = E(x, \omega)$) for this frequency range. An example of the obtained tonotopic modal profile is given in Figure \ref{tonotopic_frequency_profiles}b using a perspective view. The top view of the same profile, shown Figure \ref{tonotopic_frequency_profiles}c, is useful to graphically illustrate the characteristics of the various resonance modes. For example, the lateral mode does not appear, since it has a zero value for out-of-plane displacements, while the bending higher order modes display a larger number of local maxima compared to the lower order bending modes.

\subsection{Effects of variations in thickness or width}

\begin{figure}[H]
 \centering
 \includegraphics[scale=0.45]{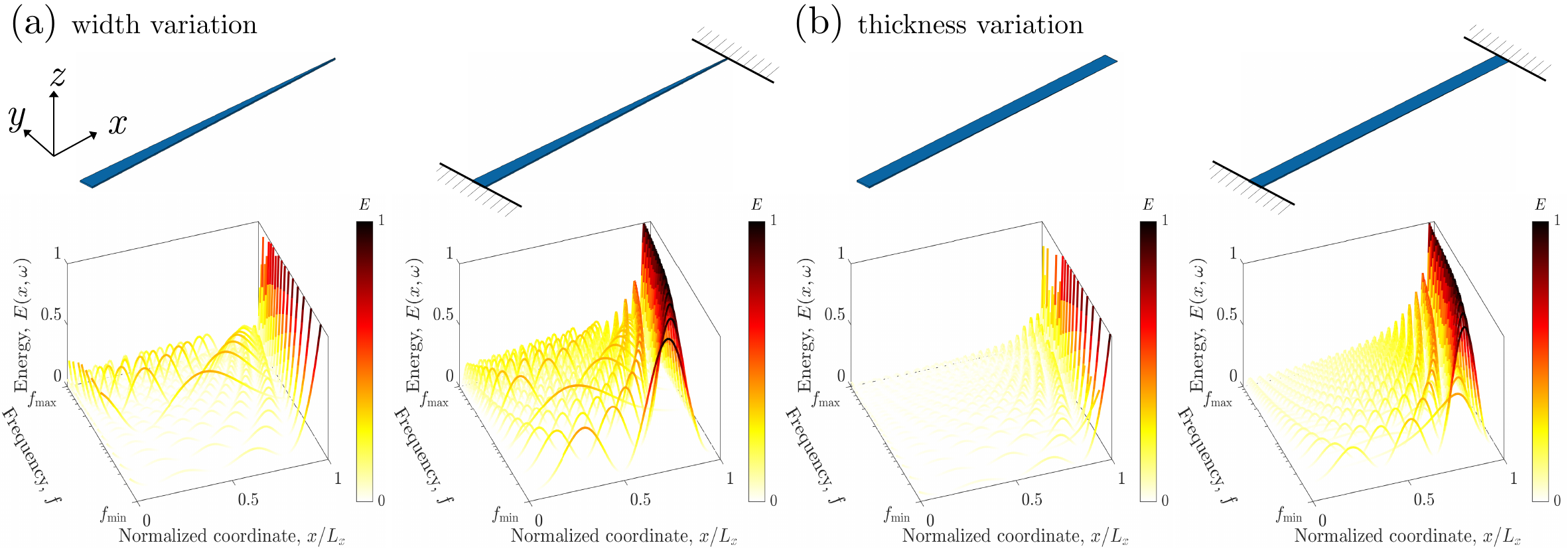}
 \caption{
 Tonotopic modal profiles for elements with variation in (a) width and (b) thickness. The distribution of mean out-of-plane displacement and its maxima for free-free and clamped-clamped boundary conditions illustrates that the symmetry of eigenmodes can be manipulated by the variation of geometric parameters and boundary conditions.
 }
 \label{example_variation_structures}
\end{figure}

The symmetry in the distribution of maxima along the element length (see Figure \ref{tonotopic_frequency_profiles}c) is due to the overall distribution of energy for each frequency. However, this symmetry can be manipulated through the control of geometric properties such as the variation of width, thickness, and the boundary conditions to which the element is subjected. To illustrate these effects, consider elements with variation in their width (Figure \ref{example_variation_structures}a) or thickness (Figures \ref{example_variation_structures}b). For each case, we display the distribution of mean displacements and their maxima for free-free and clamped-clamped boundary conditions. For instance, considering the clamped-clamped boundary conditions, creating an asymmetry in the geometric and mechanical properties of the structures allows to manipulate the distribution of energy for each mode of vibration so that maxima vary spatially according to each resonant frequency, thus inducing a tonotopic effect. The maxima, however, do not present a uniform distribution along the length of the structure, and are rather confined to a small region towards the end of the element.

In the next section, we present numerical results that demonstrate how tonotopic effects can be achieved using curved structures, additionally obtaining that the maxima of the vibration modes follow a desired spatial distribution.

\section{Tonotopic effects in curved structures} \label{results}

The numerical results presented in this section are computed considering the material properties of a polymer used in Section 3.4 for 3D-printing processes (Solflex SF650, W2P Engineering GmbH), with Young's modulus $E = 2.5$ GPa, Poisson's ratio $\nu = 0.33$, and mass density $\rho = 1150$ kg/m$^3$. The frequency range of interest is restricted to $[ f_{\min}, f_{\max} ] = [ 10^2, \, 10^4]$ Hz, thus encompassing a two-decade range. For the considered material properties and an initial thickness of $L_z = 1$ mm, the smallest flexural wavelength at $f_{\max}$ is equal to $\lambda_{\min} = ( E L_z^2 / 12 (1-\nu^2) \rho )^{1/4} ( 2\pi / f_{\max} )^{1/2} = 16.8$ mm.

\subsection{Curved profiles}


The following geometries and distribution of mean out-of-plane displacement maxima are computed as previously using the finite element method, considering isoparametric hexahedral elements with linear elastic behavior \cite{cook2001concepts,bathe1996finite,ferreira2010matlab}. An initially straight, regular node grid is mapped into a curved structure according to the geometric parameters to generate the volume described by Eqs. (\ref{geo_rz}), (\ref{geo_bh_theta}), and (\ref{geo_gap}). The $+y$ and $-y$ faces are mapped into the inner ($r_C(\theta)-b(\theta)/2$) and outer ($r_C(\theta) + b(\theta)/2$) faces, respectively, while the $-z$ and $+z$ faces are mapped into the lower ($-h(\theta)/2$) and upper ($+h(\theta)/2$) faces of the curved structure, respectively. Likewise, the $x$-coordinates of the initially straight solid element are mapped on to the $\theta$ coordinate using the linear relation $ \theta(x) = (x/L_x) \theta_{\max}$.

The boundary conditions are considered as clamped at the face with largest width of the cochlea ($\theta=0$) and free at the other end ($\theta=\theta_{\max}$), thus representing the attachment of the structure to an ideally infinitely stiff substrate. The previously presented quadratic displacement mean metric (Eq. (\ref{energy_integral})) is approximated for the use with the finite element discretization considering the sum in the radial and vertical directions, i.e.,
\begin{equation} \label{energy_sum}
 E(s,\omega) = \frac{1}{n_r n_z} \sum_r \sum_z u_z^*(s,r,z,\omega) \, u_z(s,r,z,\omega) \, ,
\end{equation}
where $n_r$ and $n_z$ are the number of nodes considered in the finite element model in the radial and vertical directions for a given $s(\theta)$ coordinate, respectively.

\begin{figure}[H]
 \centering
 \includegraphics[scale=0.45]{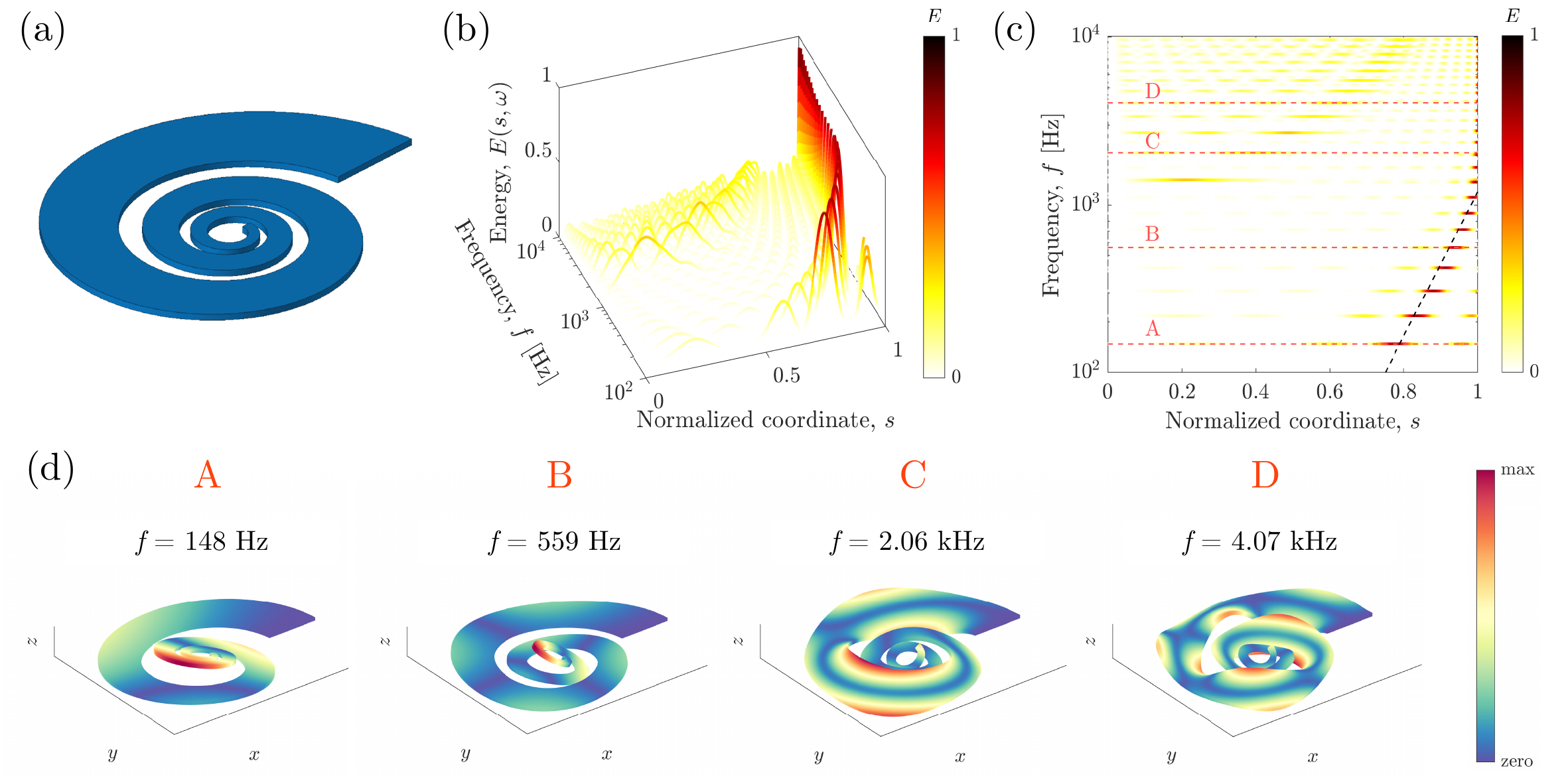}
 \caption{
 Initial investigation on the distribution of mean displacement maxima for a baseline design.
 (a) Structure obtained for $k_r = k_b = \ln(1/10)$, $k_h = 0$, $\Delta_r^{\min} = 1$ mm, and $n_T = 3$ turns.
 (b) The obtained tonotopic modal profile displays noticeable tonotopy up to $1.2$ kHz with (c) spatial separation between the peaks of the eigenmodes as their resonant frequencies increase (modes A and B); this separation is hindered at higher frequencies and includes higher order torsional modes (modes C and D). (d) Vibration modes A -- D.
 }
 \label{cochlea_baseline}
\end{figure}

A baseline structure, shown in Figure \ref{cochlea_baseline}a, is obtained by initially  considering the dimensions $L_x = 180$ mm $> 10 \lambda_{\min}$ and $L_y = 10$ mm $\approx 0.6 \lambda_{\min}$. The spiral parameters are set as $k_r = \ln(1/10)$, $k_b = \ln(1/10)$ (thus yielding a reduction to $1/10$ of the initial width), $k_h = 0$ (fixed thickness), a minimum gap of $\Delta_r^{\min} = 1$ mm, and a total of $n_T = 3$ turns. The distribution of mean displacement maxima, indicated in Figure \ref{cochlea_baseline}b, presents considerable tonotopy, with a noticeable separation between the peak for each mode up to $1.2$ kHz (as indicated by the black dashed fitting line in Figure \ref{cochlea_baseline}c), where the bending modes (A and B, Figure \ref{cochlea_baseline}d) are clearly distinguishable. Above this frequency, the maximum of each eigenmode occurs at $s = 1$, thus losing tonotopy. Also, for higher frequencies, higher-order bending modes in the radial direction (which can also be interpreted as torsional in the circumferential direction) may occur (modes C and D, Figure \ref{cochlea_baseline}d), thus hindering the desirable clear visualization of tonotopy.

\begin{figure}[H]
 \centering
 \includegraphics[scale=0.45]{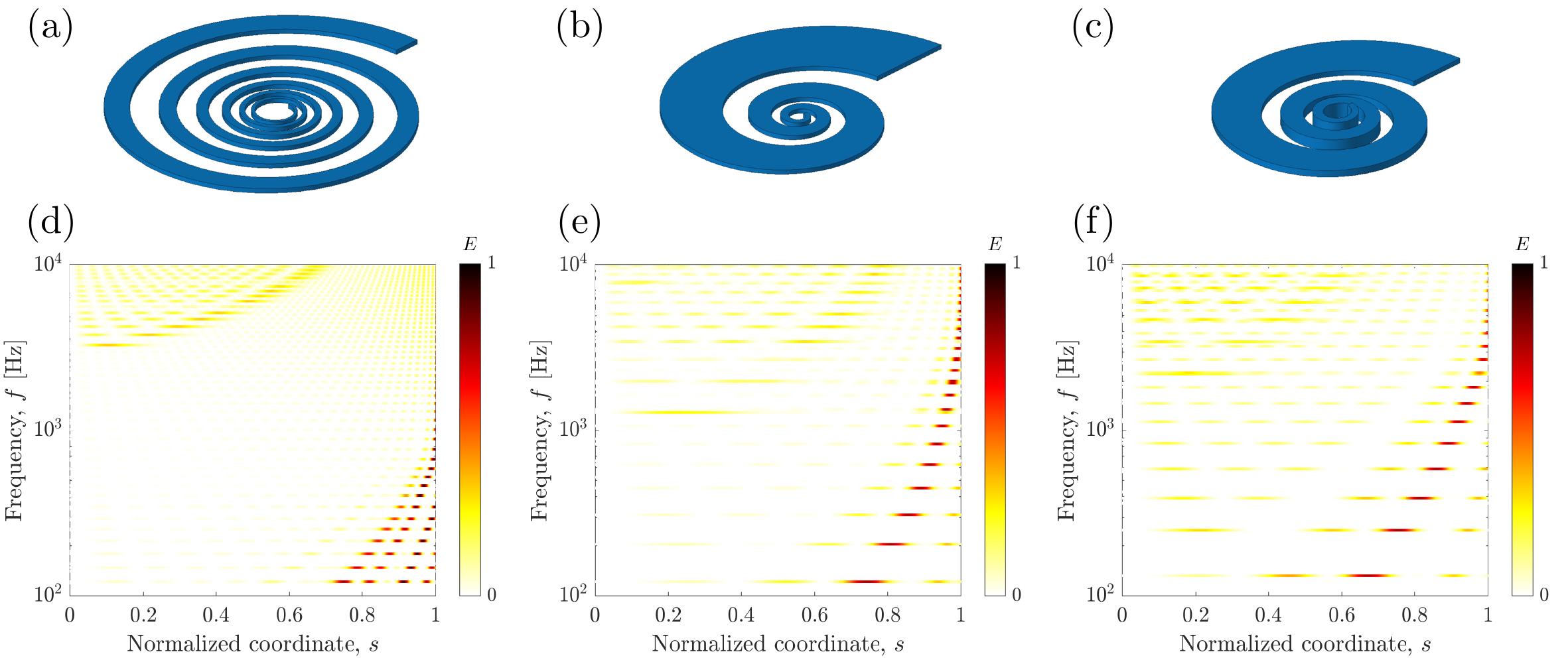}
 \caption{
 Structures obtained by increasing the (a) number of turns, (b) coiling, or (c) thickness increase of the baseline curved structure. 
 (d) The smooth change in the structure cross section yielded by the increased number of turns hinders the isolation of peaks in the distribution of mean out-of-plane displacements. 
 (e) Although limited by constraints for its physical construction, the increase in coiling facilitates the use of higher frequency modes. (f) Same observation as (e) for an increase in thickness.
 }
 \label{cochlea_parameters}
\end{figure}

To investigate the effect of the geometric parameters that describe the coiling ($k_r$, $k_b$), thickness ($k_h$), and number of turns ($n_T$) of the cochlea, three additional structures are proposed, obtained by increasing either (i) the number of turns while keeping the same final width ($n_T = 6$, Figure \ref{cochlea_parameters}a), (ii) the coiling of the spiral ($k_r = k_b = \ln(1/20)$, Figure \ref{cochlea_parameters}b), or (iii) the variation of thickness ($k_h = \ln(5)$, Figure\ref{cochlea_parameters}c), while fixing the other parameters.

The distribution of mean displacement maxima obtained for an increased number of turns, shown in Figure \ref{cochlea_parameters}d), indicates that the smooth transition in the curved structure cross section hinders the formation of isolated peaks for each resonant frequency due to a more uniform distribution of maxima along the circumferential direction for each resonant frequency.

An increase in the coiling of the structure, on the other hand, seems to be highly beneficial for a tonotopic effect in the structure, as shown in Figure \ref{cochlea_parameters}e, since the peaks corresponding to each mode show a good contrast with other maxima and also a noticeable isolation. The tuning of the structure eigenmodes through these parameters, however, may be hard to achieve in terms of manufacturing, due to the need to create extremely small parts in the resulting structure. This limitation can, however, be alleviated by an increase in the thickness of the structure, which provides similar effects as the increase in the coiling of the structure (Figure \ref{cochlea_parameters}f).

It is known that in the actual cochlea, low frequencies are mostly detected in the inner part, and high frequencies in the outer part \cite{reichenbach2014physics}. The opposite happens for elastic wave propagation in the structures considered up to this point. 
However, this is not necessarily true in all cases.
Consider, for example , a structure obtained through the same previously described process, but
stiffer at its outer edge ($s=0$) and more flexible at the inner edge ($s=1$), thus inducing, respectively, high- and low-frequency resonant frequencies in these regions.

To this end, let the geometrical control parameters be $k_r = \ln(1/5)$, $k_b = \ln(2)$, and $k_h = \ln(1/5)$. Also, all edges of the structure are considered as clamped to account for the stiffness of the surrounding medium supporting the basilar membrane in the cochlea. In this case, we also consider larger dimensions to obtain resonant frequencies comparable with the previous examples, using $L_x = 500$ mm, $L_y = 100$ mm, $L_z = 2$ mm, $\Delta_r^{\min} = 1$ mm, and $n_T = 3$ turns.

\begin{figure}[H]
 \centering
 \includegraphics[scale=0.45]{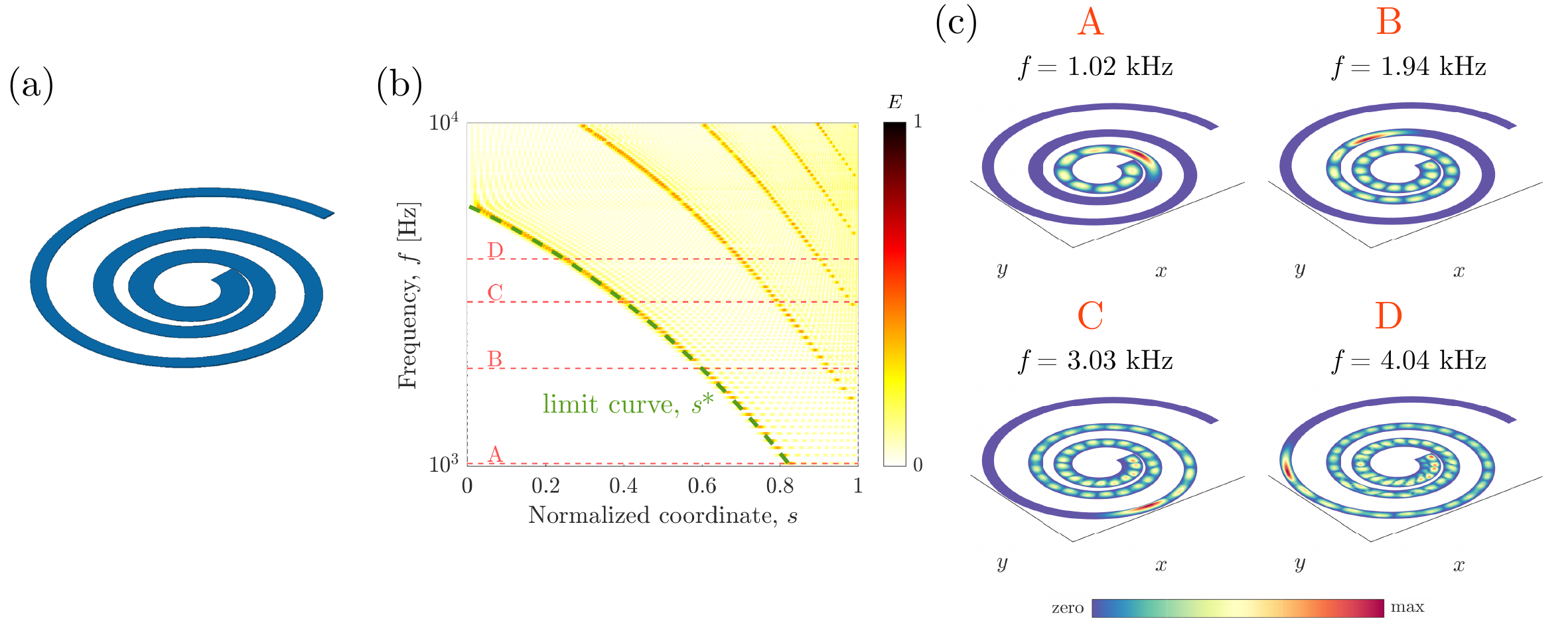}
 \caption{
 Investigation of a structure giving rise to inverted distribution of mean out-of-plane displacement maxima.
 (a) Structure obtained considering $k_r = k_h = \ln(1/5)$, $k_b = \ln(2)$, $\Delta_r^{\min} = 1$ mm, and $n_T = 3$ turns.
 (b) The obtained tonotopic modal profile shows a limit curve ($s^*$, dashed green line) confining the significant out-of-plane elastic energy to regions delimited by $s > s^*$.
 (c) Modes A--D, marked in (b), indicating a decreasing confinement of energy as the resonant frequencies increase.
 }
 \label{cochlea_inverted}
\end{figure}

These parameters result in the structure presented in Figure \ref{cochlea_inverted}a, occupying an area of approximately $567 \times 502$ mm$^2$. The associated tonotopic modal profile is shown in Figure \ref{cochlea_inverted}b, with a limit curve ($s^*$, dashed green line) indicating the location of local maxima, which confines the elastic energy at the region delimited by $s > s^*$ for each resonant frequency. Wave modes marked as A--D are displayed in Figure \ref{cochlea_inverted}c, indicating that the locations of maximum displacement move from the inner to the outer region of the cochlea as the frequency increases, as expected. For lower frequencies, the points of maxima are quite noticeable.

Although this type of structure may be interesting for certain applications and demonstrates the versatility of the proposed approach, it suffers from drawbacks such as (i) the considerable presence of void spaces in the in-plane design, thus leading to a low efficiency in the use of space, (ii) the large dimensions associated with the final structure, and (iii) the poor discrimination between maxima, which possibly hinders the use of the structure in sensing applications.

\subsection{Optimization of tonotopy}

The results from the last section show that the manipulation of tonotopy can be achieved by varying the geometric parameters that control the curvature, width, and thickness of the cochlea. The parameters that control the curvature and number of turns, however, may lead to prohibitively small features, which can be difficult in terms of manufacturing, thus suggesting that optimal values can be obtained considering practical manufacturing limitations.

\begin{figure}[H]
 \centering
 \includegraphics[scale=0.45]{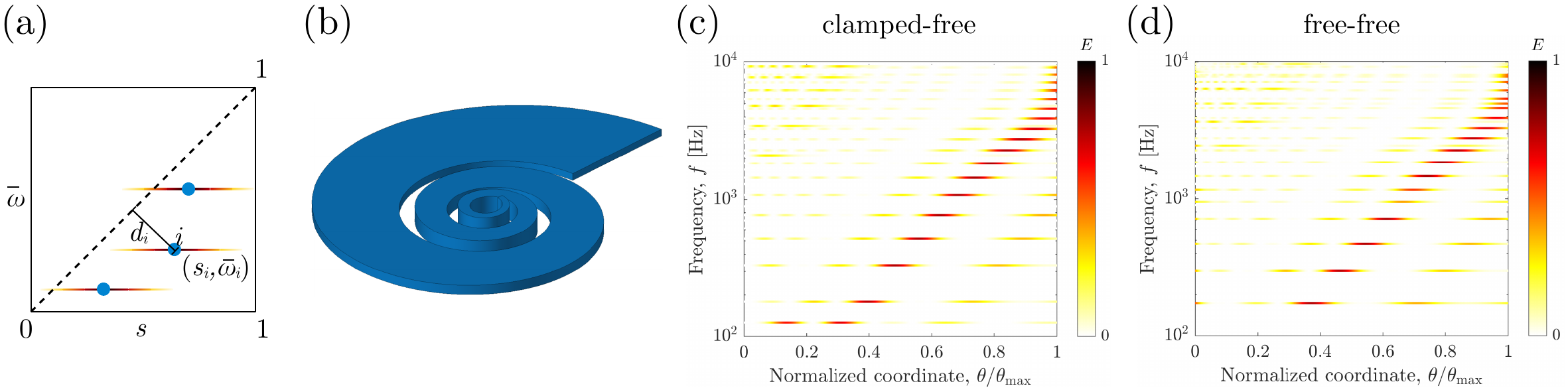}
 \caption{
 (a) Illustration of the distance considered in the quantification of tonotopy for the normalized space ($s$) and frequency ($\overline{\omega}$) coordinates, with the distance $d_i$ between the maximum of the out-of-plane mean displacements for the $i$-th mode and the $\overline{\omega} = s$ line that represents ideal tonotopy.
 (b) Obtained structure considering parameters that minimize the metric of tonotopy and (c) its tonotopic modal profile considering clamped boundary conditions at the widest radius.
 (d) Same as (c) for the free-free boundary conditions.
 }
 \label{optimized_results}
\end{figure}

To better evaluate and quantify the level of ideal tonotopy presented by the structure, we propose a metric ($\Gamma$) defined as 
\begin{equation} \label{opt_metric}
 \Gamma(k_b, \, k_r, \, k_h, \, n_T) = \sum_i d_i^2 (k_b, \, k_r, \, k_h, \, n_T) \, ,
\end{equation}
where $d_i = | s_i - \overline{\omega}_i | / \sqrt{2}$ represents the distance between the maximum of the mean displacements computed for the $i$-th out-of-plane vibration mode and the $\overline{\omega} = s$ line, as illustrated in Figure \ref{optimized_results}a, calculated considering the normalized coordinate $s_i$ and the normalized frequency $\overline{\omega}_i = \log(\omega_i/\omega_{\min}) / \log(\omega_{\max}/\omega_{\min})$, for $\omega_{\max} = 2\pi f_{\max}$ and $\omega_{\min} = 2\pi f_{\min}$. The $\overline{\omega}=s$ line is considered as the ideal distribution of modes maxima, since it covers the entire physical and considered frequency ranges with an uniform distribution. The combination of the objective function and the restrictions on variables $k_b$, $k_r$, and $k_h$ yields a constrained nonlinear optimization problem, which is solved using an active set algorithm \cite{wright1999numerical}, which can be implemented in Matlab \cite{matlaboptimization} to obtain global optimal solutions.

For the present problem, we considered $n_T = 3$ turns and a minimum gap of $\Delta_r^{\min} = 1.0$ mm (see Eq. (\ref{geo_r0})), restricting the optimization parameters to $k_r \in [ \ln(1/20), \, \ln(1/2) ]$, $k_b \in [ \ln(1/20), \, \ln(1/2) ]$, and $k_h \in [ \ln(1), \, \ln(5) ]$. The optimization results yield $k_r = \ln(1/12)$, $k_b = \ln(1/20)$, and $k_h = \ln(5)$.
Very similar results were obtained when considering normalized $\theta/\theta_{\max}$ coordinates instead of $s$ in the optimization metric given by Eq. (\ref{opt_metric}).
It is interesting to notice that $k_b$ is equal to the lower boundary of the variable, yielding a quasi-linear width decrease in the $s$-coordinate ($k_b/k_r = 1.2$, see Eq. (\ref{geo_bh_s})),  while the height increase is steeper, equal to the upper boundary of the variable ($k_h/k_r = -0.65$, see Eq. (\ref{geo_bh_s})). The resulting structure is shown in Figure \ref{optimized_results}b, and the corresponding distribution of mean displacement maxima is shown in
Figures \ref{optimized_results}c and \ref{optimized_results}d for the clamped-free and free-free boundary conditions, respectively, indicating a significant correlation when considering the normalized $\theta/\theta_{\max}$ coordinates.

\subsection{Numerical demonstration of tonotopy} \label{demonstration_tonotopy}

\begin{figure}[h!]
 \centering
 \makebox[\textwidth]{
 \includegraphics[scale=0.475]{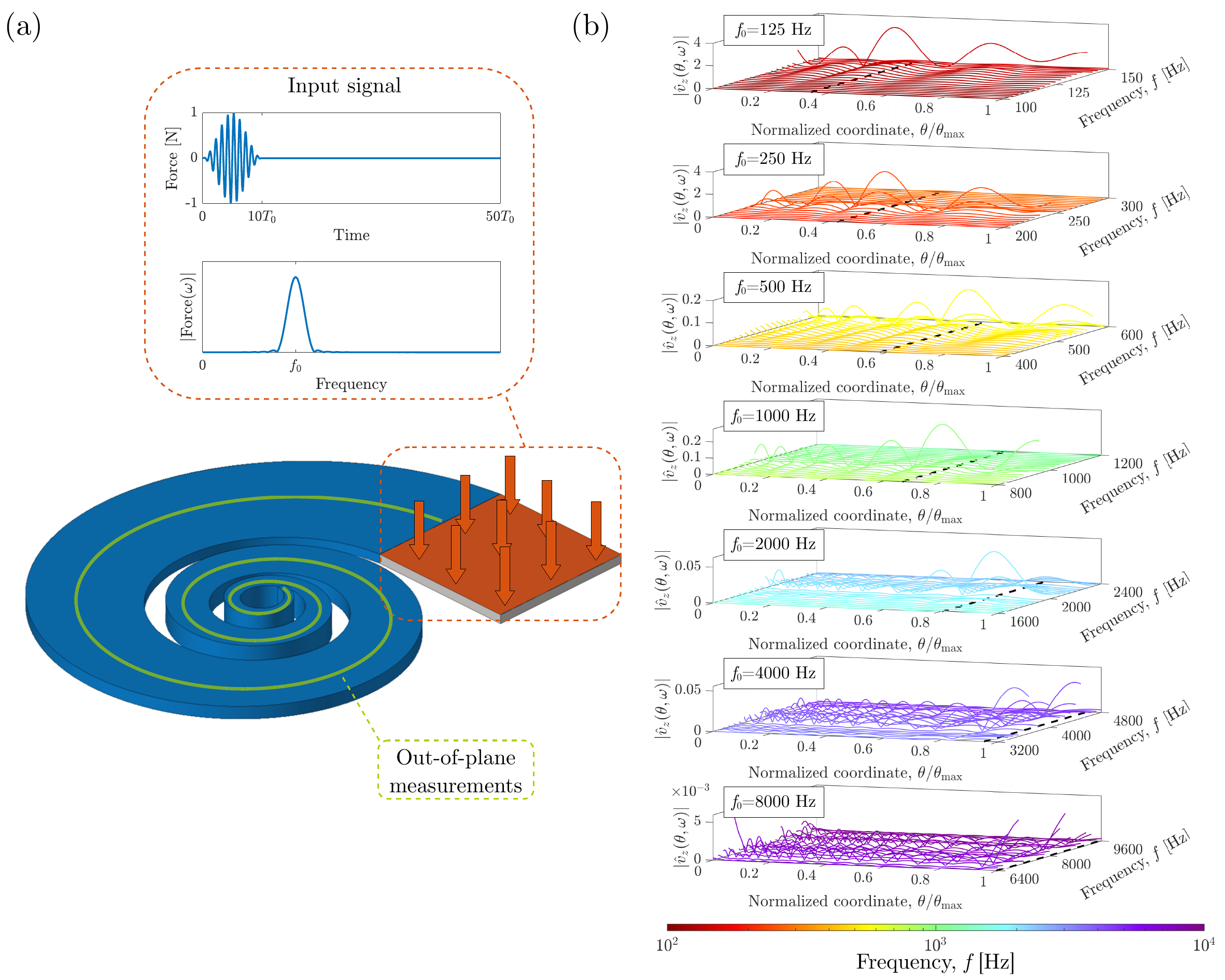}
 }
 \caption{Numerical experiment using transient signals.
    (a) The optimized spiral structure is attached to a square platform where an input force is applied on the red area, and out-of-plane velocities are computed at the cochlea center, depicted by the green line.
    Inset: example of employed burst input signal, consisting of $10$ periods contained in a Hanning window and a total of $50$ periods.
    (b) The effect of tonotopy is demonstrated by observing the varying spatial location of the maximum of the frequency components to increasing $\theta$ values as the input frequency $f_0$ increases.}
    \label{transient_frfs}
\end{figure}

The tonotopy of the resulting structure is also demonstrated by a numerical simulation where transient signals are considered. The obtained structure, shown in Figure \ref{optimized_results}b, is attached to a square platform of dimensions $16.2 \times 16.2$ mm$^2$, to which a ``burst'' signal is applied, as depicted in Figure \ref{transient_frfs}a.

For the definition of the input signal, a set of fundamental frequencies ($f_0$) is considered. For each case, the duration of the burst consists of $10$ periods, i.e., $10 T_0$, with $T_0 = 1/f_0$, while the total duration of the signal is $T = 50 T_0$ (see the inset of Figure \ref{transient_frfs}a). A total of $N_p = 2^{12}$ points is considered, thus leading to distinct values of time sampling $\Delta t = T/N_p$, which differ due to the large difference between the fundamental frequencies, chosen as $f_0 = \{ 125, \, 250, \, 500, \, 1000, \, 2000, \, 4000, \, 8000 \}$ Hz.  A time-domain transient analysis based on the Newmark method \cite{bathe1996finite} can then be performed, and the resulting out-of-plane velocity values ($v_z$) computed for the centerline of the sample (green line in Figure \ref{transient_frfs}a). A Fast Fourier Transform (FFT) is applied considering each separate $\theta$ coordinate and plotted using distinct colors for the obtained frequency components ($\hat{v}_z$), as shown in Figure \ref{transient_frfs}c.

For the first input frequency, the maximum of out-of-plane mean displacements is located close to $\theta / \theta_{\max} = 0.34$, thus resembling the results computed considering the clamped end (Figure \ref{optimized_results}c). This maximum shifts to higher values of $\theta$ as the input frequency increases, thus clearly demonstrating the effect of tonotopy. Is is also worth noticing that the resolution in the frequency domain, given by $\Delta f = 1/T$, has a smaller value for the signals with lower frequencies ($\Delta f = 2.5$ Hz for $f_0 = 125$ Hz) than for those at higher frequencies ($\Delta f = 160$ Hz for $f_0 = 8000$ Hz). An equal number of isofrequency lines are considered for each representation, thus leading to wider depicted frequency ranges for higher input frequencies, which explains the occurrence of multiple resonant frequencies for higher input frequency values, as $4000$ and $8000$ Hz.

\clearpage
\subsection{Experimental verification of tonotopy}

To verify the tonotopy of the designed cochlea and the feasibility of its potential use as a frequency selective device, we performed an experimental realization of the spiral structure
obtained by the numerical optimization process and employed in Section \ref{demonstration_tonotopy}. The structure was printed on a commercial 3D-DLP printer (Solflex SF650, W2P Engineering GmbH) using polymeric resin Solflex Tech Though (from the same manufacturer) with the same mechanical properties as considered in the numerical simulations. The printer is equipped with an UV-LED emitting at a wavelength of $385$ nm and with a specified power density of $8 \pm 0.5$ mW/cm$^2$. The nominal lateral resolution is $50$ $\mu$m and the printer supports layer thicknesses ranging from $25$ to $200$ $\mu$m. Here, the design was sliced into layers of $50$ $\mu$m, each exposed for $1.5$ s, leading to a dose of $12$ mJ/cm$^2$ per layer. The parts were oriented with their $(r,\theta)$-plane parallel to the printing platform such that the slicing was performed along the $z$-axis and
were printed standing on a removable support structure to increase the bottom surface smoothness. After printing, the parts were washed with Isopropyl Alcohol (IPA), then fully immersed in IPA and sonicated twice for $5$ minutes in an ultrasonic bath (USC$200$TH, VWR International). Finally, they were UV-post-cured for 1 hour in a PHOTOPOL Analog UV-curing unit (Dentalfarm Srl), providing all-around illumination in the wavelength range $320$--$450$ nm.

\begin{figure}[h!]
 \centering
 \makebox[\textwidth]{
 \includegraphics[scale=0.475]{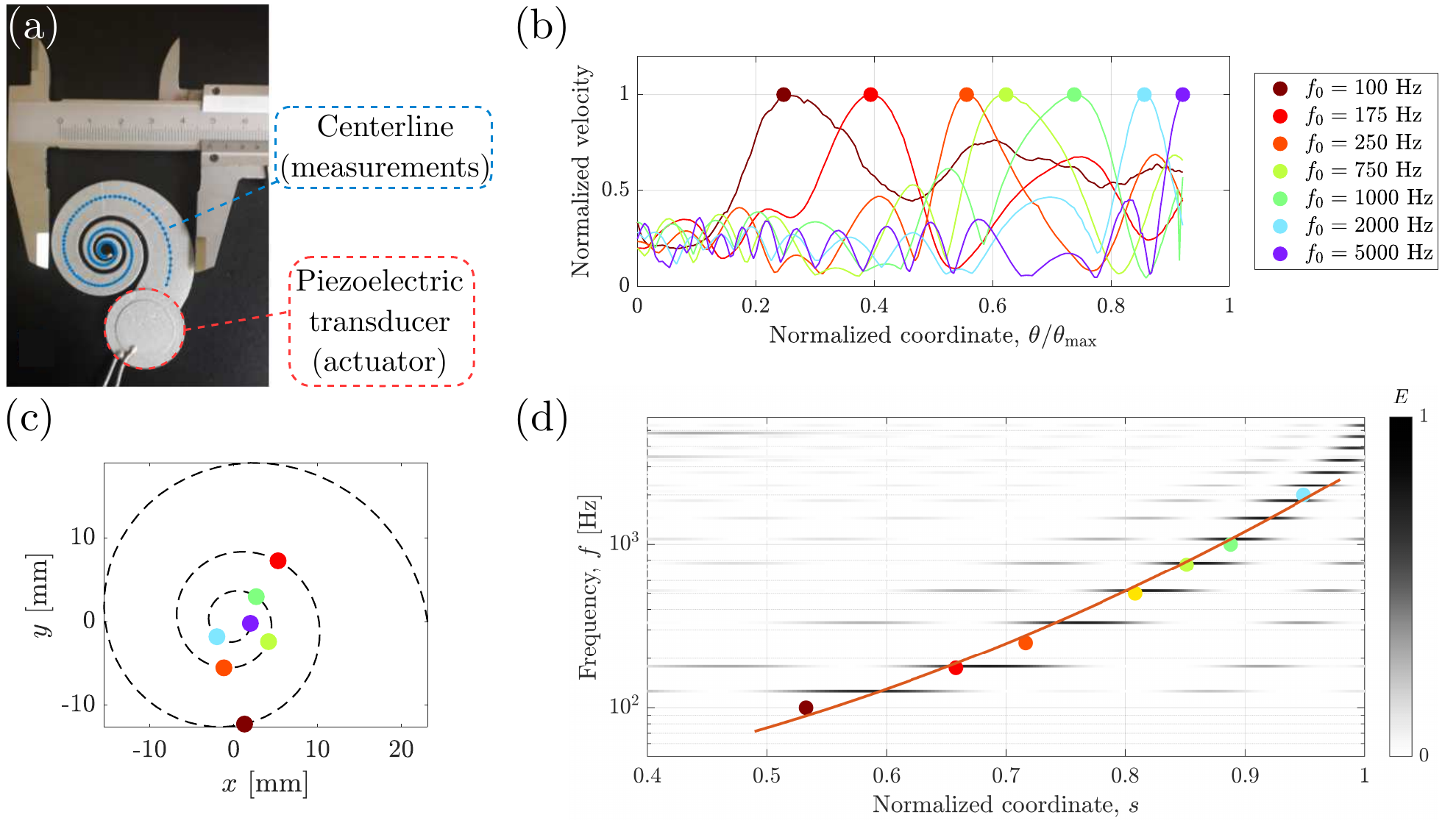}
 }
 \caption{Experimental results demonstrating the tonotopy of the considered spiral structure.
 (a) Sample and experimental set-up, showing the force input element (piezoelectric transducer, in red) and the location of the measured out-of-plane velocities along the cochlea centerline (in blue).
 (b) Measured normalized velocity amplitude detected at different points for various central frequencies ($f_0$), reported as a function of the angular position ($\theta/\theta_{\max}$) in the cochlea centerline.
 (c) Representation of each signal maximum along the measured portion of the cochlea centerline.
 (d) Distribution of the maxima along the $s$ coordinate, overlaid with the previously obtained tonotopic frequency distribution in the clamped-free configuration (Figure \ref{optimized_results}c).}
 \label{experimental_results}
\end{figure}

The experimental setup was also similar to the numerical simulation performed in Section \ref{demonstration_tonotopy}. A piezoelectric transducer, with a diameter of $d_T = 20$ mm and negligible weight with respect to the entire structure, was glued to the external edge of the printed cochlea resonator (see red circle in Figure \ref{experimental_results}a). A Gaussian shaped burst signal with varying central frequency ($f_0$) was injected with an arbitrary function generator (Agilent 33500), amplified by a factor $20$ by a liner amplifier (FLC Electronics, A400DI) and applied to the piezoelectric transducer. The central frequency $f_0$ of the signals varied in the range $\Delta f_0 = [100, \, 5000]$ Hz, with a burst duration of $T_b=10/f_0$. The experiment was repeated for each frequency $f_0$, recording the response of the structure in $200$ points along the centerline of the cochlea (see blue dots in Figure \ref{experimental_results}a), using a laser vibrometer (Polytec, OFV-505) to scan the surface of the sample. The vibrometer was placed orthogonal to the plane containing the spiral structure, in order to detect the out-of-plane velocity.

The results of the experiment are reported in Figure \ref{experimental_results}b, where the normalized amplitude of the velocity signals detected at various different points are reported as a function of the angular position on the cochlea centerline. The different coloured lines, which use the same frequency-based color coding as presented in Figure \ref{transient_frfs}, represent the responses of the structure for different values of the central frequency $f_0$ of the exciting burst and correspond, respectively, to $100$, $175$, $250$, $750$, $1000$, $2000$, and $5000$ Hz. Results show that the maximum (highlighted by a colored dot) moves along the cochlea centerline starting from the outer part (maximum radius) to the apex (minimum radius), where higher frequencies are detected. The maxima are well resolved in space, as highlighted by reporting the position of the maximum amplitude on the cochlea centerline (see Figure \ref{experimental_results}c), and each maximum is unique for a fixed frequency along the cochlea, thus proving the tononotopy of the structure.

By plotting the logarithm of the central frequency as a function of the position of the maximum in the normalized $s$ coordinate for a given set of central frequencies, it is possible to recognize a clear correlation with the tonotopic modal profile previously presented in Figure \ref{optimized_results}c; for greater clarity, results are overlaid in Figure \ref{experimental_results}d. Also, a functional dependence between the central frequency and its location in the cochlea centerline can be estimated by an exponential fit of the form $\tilde{f}(s) = g e^{k \, l(\theta)^2}$ (see Eq. (\ref{s})), with $g=22$ Hz and $k=0.7$ m$^{-2}$. In this case, the frequency range is reduced to $[100, \, 2000]$ Hz to allow for single fit, instead of adopting a piecewise function. Also, an additional measurement is performed at $f_0 = 500$ Hz (yellow point in Figure \ref{experimental_results}c) to confirm its closeness to the exponential fit. This fit may allow to identify the optimal position for the detection on the structure of a certain frequency in the range $\Delta f_0$.

\clearpage
\subsection{Sensing applications}

\begin{figure}[h!]
 \centering
 \makebox[\textwidth]{
 \includegraphics[scale=0.475]{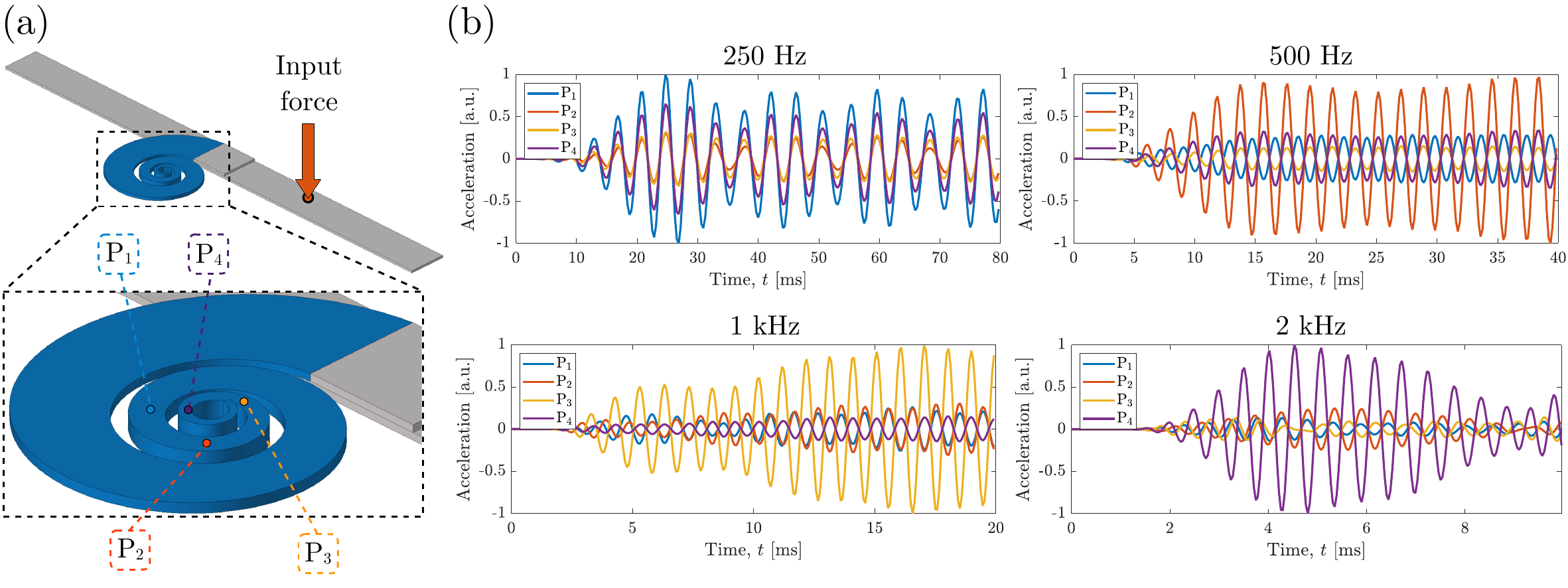}
 }
 \caption{Sensing application for the cochlea-inspired resonator as a frequency detection device of waves traveling in a beam.
 (a) An input force is applied at the upper face of the beam, with output points on the cochlea labeled as P$_1$, P$_2$, P$_3$, and P$_4$, with increasing $\theta$ coordinates.
 (b) Computed out-of-plane accelerations, normalized with respect to the the largest acceleration value for each input frequency: $250$ Hz, $500$ Hz, $1$ kHz, and $2$ kHz.
 Maximum corresponding amplitudes are detected at P$_1$, P$_2$, P$_3$, and P$_4$, respectively.
 }
 \label{application_pure_sensing}
\end{figure}

Having assessed the validity of the tonotopic effect on the optimized spiral structure, we now propose a numerical experiment to demonstrate its possible use in sensing and in non-destructive testing applications. For this purpose, a $216 \times 16 \times 1$ mm$^3$ beam is attached to the previously designed cochlea and platform structure. The beam is then excited with an out-of-plane point force at $40$ mm from the edge of the platform, located at its upper face and applied at its center. Points $P_1$, $P_2$, $P_3$, and $P_4$, corresponding to the frequencies of $250$ Hz, $500$ Hz, $1$ kHz, and $2$ kHz are chosen as outputs. As these frequencies are not necessarily resonant frequencies, the location of the points at the central line of the cochlea are determined by a linear interpolation of the peaks presented in Figure \ref{optimized_results}c for the $\theta$ coordinate, resulting in the angles $\theta/2\pi = 1.3$, $\theta/2\pi = 1.6$, $\theta/2\pi = 2.0$, and $\theta/2\pi = 2.4$, respectively. The resulting structure is presented in Figure \ref{application_pure_sensing}a. Distinct input forces with single frequency components are applied for a total of $10$ periods (using a discretization of $2^8$ points in the time domain) and a Hanning window (see the inset in Figure \ref{transient_frfs}a), while the output signals are computed for a total of $20$ periods. The normalized acceleration output values are presented in Figure \ref{application_pure_sensing}b. In each case, the single-frequency input signal generates the strongest response at the corresponding output point, i.e., $250$ Hz for point P$_1$, $500$ Hz for point P$_2$, $1$ kHz for point P$_3$, and $2$ kHz for point P$_4$. Furthermore, although excitation signals at lower frequencies still produce a considerable response at the high-frequency output points (e.g., the output at P$_4$ is appreciable even for the $250$ Hz input signal), this effect is much less pronounced at higher frequencies. In particular, for an input frequency of $2$ kHz, the output accelerations at points P$_1$ -- P$_3$ are considerably smaller than at P$_4$, which can be explained by the localization of out-of-plane mean displacements of higher frequency wave modes.

\begin{figure}[h!]
 \centering
 \makebox[\textwidth]{
 \includegraphics[scale=0.475]{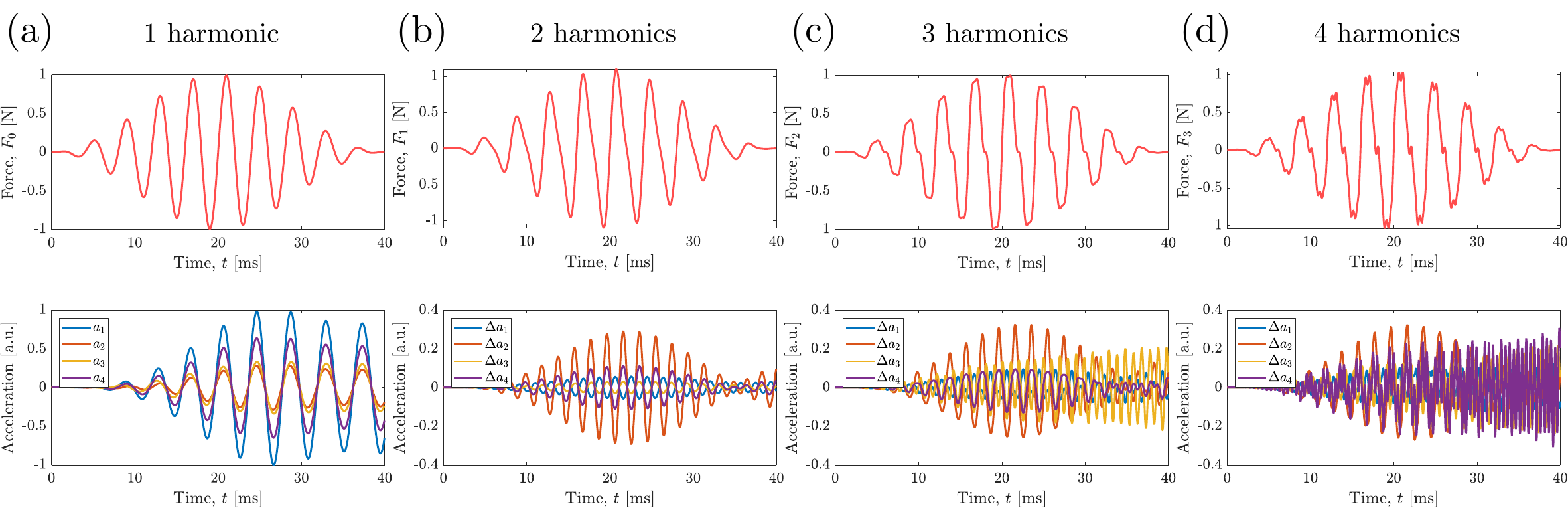}
 }
 \caption{
 Input force (upper row) and normalized acceleration outputs (lower row)
 considering input signals with multiple harmonics, namely
 (a) $250$ Hz,
 (b) $250$ and $500$ Hz,
 (c) $0.25$, $0.5$, and $1$ kHz, and
 (d) $0.25$, $0.5$, $1$, and $2$ kHz.
 }
 \label{application_mixed_sensing}
\end{figure}

\clearpage
As a second application, let us now consider the problem of determining the frequency content of a signal ($F_n(t)$) containing multiple harmonics, which may derive, for instance, from nonlinearities. Let us define such a signal as $F_n(t) = \sum_{i=0}^{n} w_i \sin(2^i \omega_0 t)$, for $\omega_0 = 2\pi f_0$, which is applied as the input force of the structure shown in Figure \ref{application_pure_sensing}a, considering the fundamental frequency $f_0 = 250$ Hz and $3$ additional harmonics ($500$ Hz, $1$ kHz, and $2$ kHz). For the sake of numerical evaluation, we consider $w_0=1$, $w_1 = 1/2^2$, $w_2 = 1/2^3$, and $w_3 = 1/2^4$. The representation of signals $F_0$, $F_1$, $F_2$, and $F_3$, applied for a total of $10$ periods relative to the $f_0$ frequency using $2^{10}$ points are shown in the first row of Figure \ref{application_mixed_sensing}.

The normalized acceleration outputs at points P$_1$ -- P$_4$, computed considering the single frequency signal $F_0$ and labeled respectively as $a_1$ -- $a_4$, are shown in Figure \ref{application_mixed_sensing}a, where all output signals oscillate with the same frequency. As the number of harmonics is increased, a differential acceleration quantity, $\Delta a_i$, can be computed at the $i$-th point as $\Delta a_i = a_i - a_i^{\text{ref}}$, where $a_i^{\text{ref}}$ is taken as the reference case shown in Figure \ref{application_mixed_sensing}a. The corresponding variation for each differential acceleration is shown in Figures \ref{application_mixed_sensing}b--d for the inputs $F_1$ -- $F_3$, respectively, where it is possible to notice a variation in the differential acceleration of each point as the number of harmonics used to compose the excitation signal increases.

The differential acceleration can also be used to compute the integral $I_i = \int \Delta a_i^2 \, dt$, thus indicating an increase in the acceleration associated to the $i$-th point. The results are indicated in Table \ref{table_I} for each distinct excitation, where it may be noticed that (i) for 2 harmonics ($F_1$), $I_2$ is clearly larger than the other quantities, which is also indicated in Figure \ref{application_mixed_sensing}b; (ii) for 3 harmonics ($F_2$), $I_3$ presents a substantial increase; (iii) for 4 harmonics ($F_3$), $I_4$ presents the most noticeable increase. Thus, it is possible to demonstrate that an increase in the number of harmonics can be detected by each of the corresponding output points through a metric associated to the measured acceleration.

\begin{table}[h!]
 \caption{Integral over time for square of differential accelerations.}
 \centering
 \begin{tabular}{c|ccc}
 \hline
   & \multicolumn{3}{c}{Excitation} \\
  $I_i$ [a.u.] & $F_1$ (2 harm.) & $F_2$ (3 harm.) & $F_3$ (4 harm.) \\
  \hline
  $I_1 \times 10^{-3}$ & $0.0279$ & $0.0539$ & $0.0767$ \\
  $I_2 \times 10^{-3}$ & $0.5691$ & $0.5868$ & $0.5935$ \\
  $I_3 \times 10^{-3}$ & $0.0076$ & $0.3460$ & $0.3586$ \\
  $I_4 \times 10^{-3}$ & $0.0840$ & $0.0894$ & $0.5411$ \\
  \hline
 \end{tabular}
 \label{table_I}
\end{table}

\clearpage
\section{Conclusions} \label{conclusions}

In summary, we have proposed a new efficient design for cochlea-inspired tonotopic materials, created by sweeping a rectangular area along a logarithmic spiral curve. The resulting structure presents a distribution of out-of-plane displacement maxima along its centerline, which can be modified by exploiting the geometrical parameters that control the curvature, width, and height of the structure, thus achieving a tunable tonotopic effect. Although torsional modes (in the radial direction) can occur for higher frequencies, bending modes (in the circumferential direction) dominate the vibration modes spectrum.

The resulting tonotopy can be optimized by distributing the modal displacements maxima along a linear spatial distribution that realizes ideal tonotopy, i.e., uniformly distributed resonant frequencies along the structure. Using this procedure, we were able to design and fabricate a structure which displays a uniform distribution of mean displacement maxima over nearly two decades, as demonstrated both numerically and experimentally. The potential use of such structure in sensing applications considering output channels located along its centerline was also numerically demonstrated for signals with single and multiple frequency components.

The presented results open new possibilities, through targeted designs, for applications of tonotopic bioinspired materials in energy harvesting, non-destructive testing, and vibration attenuation.

\section*{Acknowledgments}

VFDP, FB, NMP, and ASG  are  supported  by  the EU H2020  FET  Open ``Boheme''  grant  No. 863179. JT and DU are supported by acknowledge the Norwegian University of Science and Technology for funding the PhD fellowship under project number 81148180.

\section*{Author contributions}

ASG conceived the manuscript idea;
VFDP performed the numerical analyses and simulations;
DU and JT performed the manufacturing of samples;
ASG performed the experiments on manufactured samples;
NMP and ASG supervised the work;
VFDP, FB, and ASG wrote the first version of the manuscript.
All authors contributed to the discussion of the manuscript and gave approval to the final version.

\section*{Competing interests}

The authors declare no competing interests.

\bibliographystyle{elsarticle-num}
\bibliography{cochlea_resonators}

\end{document}